\pdfoutput=1

\documentclass[11pt,twoside,a4paper,cmspaper,final,collab]{cms-tdr}
\def\svnVersion{403836}\def\cmsCernNoTag{CERN-EP/2016-258}\def\cmsCernDate{\today}\def\cmsMessage{Published in Physical Review Letters as \href{http://dx.doi.org/10.1103/PhysRevLett.118.162301}{\doi{10.1103/PhysRevLett.118.162301.}}}

\begin{document}\cmsNoteHeader{HIN-16-004}

\hyphenation{had-ron-i-za-tion}
\hyphenation{cal-or-i-me-ter}
\hyphenation{de-vices}
\RCS$Revision: 391642 $
\RCS$HeadURL: svn+ssh://svn.cern.ch/reps/tdr2/papers/HIN-16-004/trunk/HIN-16-004.tex $
\RCS$Id: HIN-16-004.tex 391642 2017-03-05 20:14:31Z alverson $
\newlength\cmsFigWidth
\ifthenelse{\boolean{cms@external}}{\setlength\cmsFigWidth{0.85\columnwidth}}{\setlength\cmsFigWidth{0.4\textwidth}}
\ifthenelse{\boolean{cms@external}}{\providecommand{\cmsLeft}{top\xspace}}{\providecommand{\cmsLeft}{left\xspace}}
\ifthenelse{\boolean{cms@external}}{\providecommand{\cmsRight}{bottom\xspace}}{\providecommand{\cmsRight}{right\xspace}}
\ifthenelse{\boolean{cms@external}}{\providecommand{\CL}{C.L.\xspace}}{\providecommand{\CL}{CL\xspace}}
\ifthenelse{\boolean{cms@external}}
{\providecommand{\suppMaterial}{the supplemental material [URL will be inserted by publisher]}}
{\providecommand{\suppMaterial}{Appendix~\ref{app:suppMat}}}
\newcommand{\mumu}{\ensuremath{\PGmp\PGmm}\xspace}
\newcommand{\Jpsi}{\JPsi\xspace}
\newcommand{\psiP}{\ensuremath{\psi\mathrm{(2S)}}\xspace}
\newcommand{\doubleRatio}{\ensuremath{(N_{\psiP}/ \allowbreak N_{\Jpsi})_{\PbPb} / \allowbreak (N_{\psiP}/ \allowbreak N_{\Jpsi})_{\pp}}\xspace}
\newcommand {\npart}  {\ensuremath{N_{\text{part}}}\xspace}
\newcommand{\pp}{\ensuremath{\Pp\Pp}\xspace}
\newcommand{\pPb}{\ensuremath{\Pp\mathrm{Pb}}\xspace}
\newcommand{\PbPb}{\ensuremath{\mathrm{PbPb}}\xspace}
\newcommand{\AuAu}{\ensuremath{\mathrm{AuAu}}\xspace}
\newcommand{\sqrtsnn}{\ensuremath{\sqrt{s_{_{\mathrm{NN}}}}}\xspace}
\newcommand{\Lxyz}{\ensuremath{L_{xyz}}\xspace}
\newcommand{\ctauxyz}{\ensuremath{\ell_{\Jpsi}^\mathrm{3D}}\xspace}

\hyphenation{pa-ra-me-ters}
\cmsNoteHeader{HIN-16-004}
\title{\texorpdfstring{Relative modification of prompt \psiP and \Jpsi yields from \pp to \PbPb collisions at $\sqrtsnn = 5.02$\TeV}{Relative modification of prompt psi(2S) and J/psi yields from pp to \PbPb collisions at sqrt(s[NN]) = 5.02 TeV}}

\date{\today}

\abstract{
The relative modification of the prompt \psiP and \Jpsi yields from pp to \PbPb collisions, at the center of mass energy of 5.02\TeV per nucleon pair, is presented.
The analysis is based on \pp and \PbPb data samples collected by
the CMS experiment at the LHC in 2015, corresponding to integrated luminosities of 28.0\pbinv and 464\mubinv, respectively. The double ratio of measured yields of prompt charmonia reconstructed through their decays into muon pairs, $(N_{\psiP}/N_{\Jpsi})_{\PbPb}/ (N_{\psiP}/N_{\Jpsi})_{\pp}$, is determined as a function of \PbPb collision centrality and charmonium transverse momentum \pt, in two kinematic intervals: $\abs{y} < 1.6$ covering $6.5 < \pt < 30\GeVc$ and $1.6 < \abs{y} < 2.4$ covering $3 <\pt< 30\GeVc$. The centrality-integrated double ratios are $0.36 \pm 0.08\stat\pm0.05\syst$  in the first interval and $0.24 \pm 0.22\stat \pm 0.09\syst$ in the second. The double ratio is lower than unity in all the measured bins, suggesting that the \psiP yield is more suppressed than the \Jpsi
yield in the explored phase space.
}

\hypersetup{%
pdfauthor={CMS Collaboration},%
pdftitle={Relative modification of prompt psi(2S) and J/psi yields from pp to PbPb collisions at sqrt(s[NN]) = 5.02 TeV},%
pdfsubject={CMS},%
pdfkeywords={CMS, physics, heavy ions, quarkonia, J/psi, psi(2S), \PbPb, 5.02 TeV}}

\maketitle

Quarkonium production is expected to be significantly influenced by the formation of a
quark-gluon plasma (QGP) in heavy ion collisions, thereby providing an important
probe of the QGP properties.
While the early-formed mesons propagate through the medium and probe its space-time evolution,
the overall production rates can also reflect later production mechanisms.
The suppression of charmonium production due to Debye screening of the color charges in the plasma was proposed 30 years ago~\cite{Matsui:1986dk}. The  \JPsi suppression observed in \PbPb collisions at the SPS by NA50~\cite{Alessandro:2004ap} and
in AuAu collisions at RHIC by PHENIX~\cite{Adare:2006ns} is compatible with this picture. Another effect, referred to as regeneration, might be at work at sufficiently high collision energy, when the number of charm-anticharm pairs is large: uncorrelated charm quarks and antiquarks may coalesce in the medium to form a bound charmonium state, leading to an enhanced production in heavy ion collisions~\cite{BraunMunzinger:2000px,Thews:2000rj}. Hints of the latter were found at the LHC in recent results from ALICE~\cite{Abelev:2013ila,Adam:2016rdg}, which measured a weaker \JPsi meson suppression than at RHIC,
especially at low \pt.

The study of the modification of the excited \psiP state is of particular interest. The strength of medium effects on its production might be significantly different from that of the \JPsi because of the larger size and weaker binding of the \psiP state. The smaller binding energy should make it easier for the \psiP to dissociate in the medium, leading to sequential melting~\cite{Digal:2001ue}. However, the smaller production cross section and branching fraction to dimuons make the \psiP less accessible experimentally than the
\JPsi,
especially when a large background is present, such as in heavy ion collisions. At the SPS fixed-target facility, the \psiP production in heavy ion collisions was seen to be more suppressed than the \JPsi by NA38~\cite{Baglin:1994ui},
NA50~\cite{Alessandro:2006ju}, and NA60~\cite{Arnaldi:2007zz}, in SU, PbPb, and InIn collisions, respectively.

A useful variable to compare the strength of medium effects on the
\JPsi and \psiP in \PbPb collisions is the double ratio \doubleRatio,
which is the
ratio of the corresponding nuclear modification factors. While Debye screening in the hot medium should make the double ratio smaller than unity, the presence of regeneration effects could make it exceed unity, if uncorrelated quark coalescence produces \psiP mesons more frequently than \JPsi mesons.
The double ratio allows
for the partial to total cancellation of corrections (including acceptance, efficiency, and integrated luminosity) and their associated uncertainties. The CMS measurement of
the prompt charmonium double ratio
at a center-of-mass energy per nucleon pair of $\sqrtsnn = 2.76\TeV$~\cite{hin-12-007} showed that the \psiP
is more suppressed than the \JPsi
at midrapidity and high transverse momentum ($\abs{y}<1.6$, $6.5 < \pt < 30\GeVc$), while
at more forward rapidity and intermediate \pt ($1.6<\abs{y}<2.4$, $3 <\pt < 30\GeVc$), a smaller suppression of the \psiP than the \Jpsi was favored. This behavior could be reproduced by  introducing a different
time dependence of the \JPsi and \psiP regeneration processes~\cite{Du:2015wha}, or by considering different possible heavy quark potentials~\cite{Chen:2016vha}. A similar measurement from the ALICE experiment~\cite{Adam:2015isa}, integrated over \pt and at forward rapidity ($2.5<y<4$), favored the \psiP
to be more suppressed than the
\JPsi, as expected in other models~\cite{Andronic:2009sv,Chen:2013wmr}.
The medium effects (Debye screening, regeneration, and others) affecting the two charmonia might have different dependences on the collision energy, emphasizing the relevance of performing measurements at several energies.

In this Letter, we report a new study of \JPsi and \psiP
relative production in \pp and \PbPb data collected with the CMS experiment at the CERN LHC in 2015, at $\sqrtsnn = 5.02\TeV$. The larger integrated luminosities allow for a more precise and differential measurement of the double ratio
as a function of centrality and, for the first time, as a function of the charmonium \pt.

The central feature of the CMS apparatus is a superconducting solenoid of 6\unit{m} internal diameter, providing a magnetic field of 3.8\unit{T}. Within the solenoid volume are a silicon pixel and strip tracker, a lead tungstate crystal electromagnetic calorimeter, and a brass and scintillator hadron calorimeter, each composed of a barrel and two endcap sections. Forward calorimeters extend the coverage provided by the barrel and endcap detectors. Muons are measured in the pseudorapidity range $\abs{\eta} < 2.4$ in gas-ionization detectors embedded in the steel flux-return yoke outside the solenoid, with detection planes made using three technologies: drift tubes, cathode strip chambers, and resistive plate chambers. Matching muons to tracks measured in the silicon tracker leads to a relative transverse momentum resolution
between 1 and 2\% for a typical muon in this analysis ($\pt< 30\GeVc$)~\cite{Chatrchyan:2012xi}. A more detailed description of the CMS detector, together with a definition of the coordinate system used and the relevant kinematic variables, can be found in Ref.~\cite{Chatrchyan:2008zzk}.

Hadronic collisions are selected using information from the forward hadron calorimeters (HF), covering $2.9<\abs{\eta}<5.2$, in coincidence with a bunch crossing identified by beam pick-up timing detectors. A primary vertex reconstructed with at least two tracks is also required. In addition, a filter
is applied
on the compatibility of the silicon pixel cluster width distribution
and the vertex position.
For \PbPb collisions only, at least three towers above 3\GeV are requested in the HF on each side of the interaction point. Centrality is defined using fractions of the inelastic hadronic cross section determined from the HF distributions, with 0\% denoting the most central
collisions~\cite{Chatrchyan:2011pb}.

The integrated luminosities are 28.0\pbinv for \pp data and 464\mubinv for \PbPb data. The dimuon ratios reported in this paper are unaffected by the small number of extra collisions potentially present in the collected events: the mean of the Poisson distribution of the number of collisions per bunch crossing (pileup),
averaged over the full data sample, is approximately 0.9 for the \pp data and much smaller for the \PbPb data.
Dimuon events are selected by the level-1 trigger system, with no explicit muon momentum threshold. The 0--30\% most central events have a prescale needed to reduce their high trigger rates, corresponding to an effective integrated luminosity of 351\mubinv.

Simulated events are used to tune the muon selection criteria and the signal fitting parameters, as well as for acceptance and efficiency studies. These Monte Carlo (MC) samples, produced using \PYTHIA~8.209~\cite{Sjostrand:2008aa}, are embedded in a realistic \PbPb background event generated with \textsc{hydjet 1.9}~\cite{Lokhtin:2005px} and propagated through the CMS detector with \GEANTfour~\cite{Agostinelli:2002hh}. These events are processed through the trigger emulation and the event reconstruction
chain.

The muon reconstruction algorithm starts by finding tracks in the muon detectors, which are then fitted together with tracks reconstructed in the silicon tracker. Kinematic limits are imposed on
the single muons so that their reconstruction efficiency stays above 10\%.
These limits are $\pt^{\PGm}>3.5\GeVc$ for $\abs{\eta^{\PGm}}<1.2$, $\pt^{\PGm}>1.8\GeVc$ for $2.1<\abs{\eta^{\PGm}}<2.4$, and linearly interpolated in the intermediate $\abs{\eta^{\PGm}}$ region. The muons are required to match those used online by the dimuon trigger, to be of opposite charge,
and to survive standard quality selection criteria~\cite{Chatrchyan:2012xi}.
In order to remove cosmic-ray muons, the transverse and longitudinal distances of closest approach between the muon trajectory and the reconstructed primary vertex are required to be less than 0.3\unit{cm} and 20\unit{cm}, respectively. 
The fit probability that the two muon tracks originate from a common vertex is required to be larger than 1\%.

Nonprompt charmonia,
originating from the decays of B mesons, are resolved using  the pseudo-proper decay length $\ctauxyz = c \,\Lxyz \, m_{\JPsi}/|p_{\mu\mu}|$,
where $\Lxyz$ is the distance between the primary and dimuon vertices, $m_{\JPsi}$ the mass of the \JPsi meson (assumed for all dimuon candidates), and $p_{\mu\mu}$ the dimuon momentum. Dimuons are discarded if their $\ctauxyz$ is larger than a $l_{0}$ threshold, computed using MC simulations to keep 90\% of the prompt \JPsi.
Since the \ctauxyz resolution improves with increasing dimuon \pt, from ${\approx}100\mum$ to ${\approx}20\mum$ in this analysis, the $l_{0}$ cut values also depend on \pt.
This selection removes more than 80\% of the nonprompt \JPsi.
The double ratio of prompt charmonia is deduced from the double ratio of charmonia passing the \ctauxyz selection.
This is accomplished taking into account the \ctauxyz selection efficiencies for prompt ($\epsilon_P$) and nonprompt ($\epsilon_{NP}$) charmonia,
both estimated from simulation studies.
The contamination from nonprompt charmonia is also accounted for, using dimuons failing the \ctauxyz selection:
$f_P = (f_\text{pass} - \epsilon_{NP}) / (\epsilon_P - \epsilon_{NP})$, with $f_P$ the fraction of prompt charmonia and $f_\text{pass}$ the fraction of charmonia passing the \ctauxyz selection.
This correction changes the double ratio by values that depend on the analysis bin but are always smaller than 0.09.

The \psiP to \JPsi yield ratios, $N_{\psiP}/N_{\JPsi}$, are extracted
in \pp and \PbPb collisions from unbinned maximum extended likelihood fits of the \mumu invariant mass distributions in the region $2.2 < m_{\mumu} < 4.5\GeVc^{2}$. The analysis is carried out differentially in charmonium \pt and event centrality, as well as integrated over these variables, for two kinematic ranges:
$\abs{y}<1.6$, $6.5 < \pt < 30 \GeVc$ and
$1.6<\abs{y}<2.4$, $3 < \pt < 30 \GeVc$. The different lower \pt thresholds reflect the detector acceptance.

In the fit of the \pp dimuon mass distribution, the \JPsi resonance is described by two Crystal Ball (CB) functions~\cite{SLAC-R-236}, 
with common mean and tail parameters but independent widths and free relative amplitudes (seven free parameters).
In the \PbPb case, the CB tail parameters and the ratio between the widths of the two CB functions are fixed to the values extracted from simulation studies.
In both cases, the shape of the \psiP is determined by the shape of the \JPsi, all parameters being identical except for the mean and width, 
which are scaled by the \psiP over \JPsi mass ratio.
The background is described by a polynomial of order $N$, where $N$ is the lowest value that provides a good description of the data and is determined in each analysis bin by performing a log-likelihood ratio (LLR) test between polynomials of different orders, while keeping the signal parameters fixed; it is never larger than 3.

\begin{figure}[htbp]
 \centering
    \includegraphics[width=0.49\textwidth]{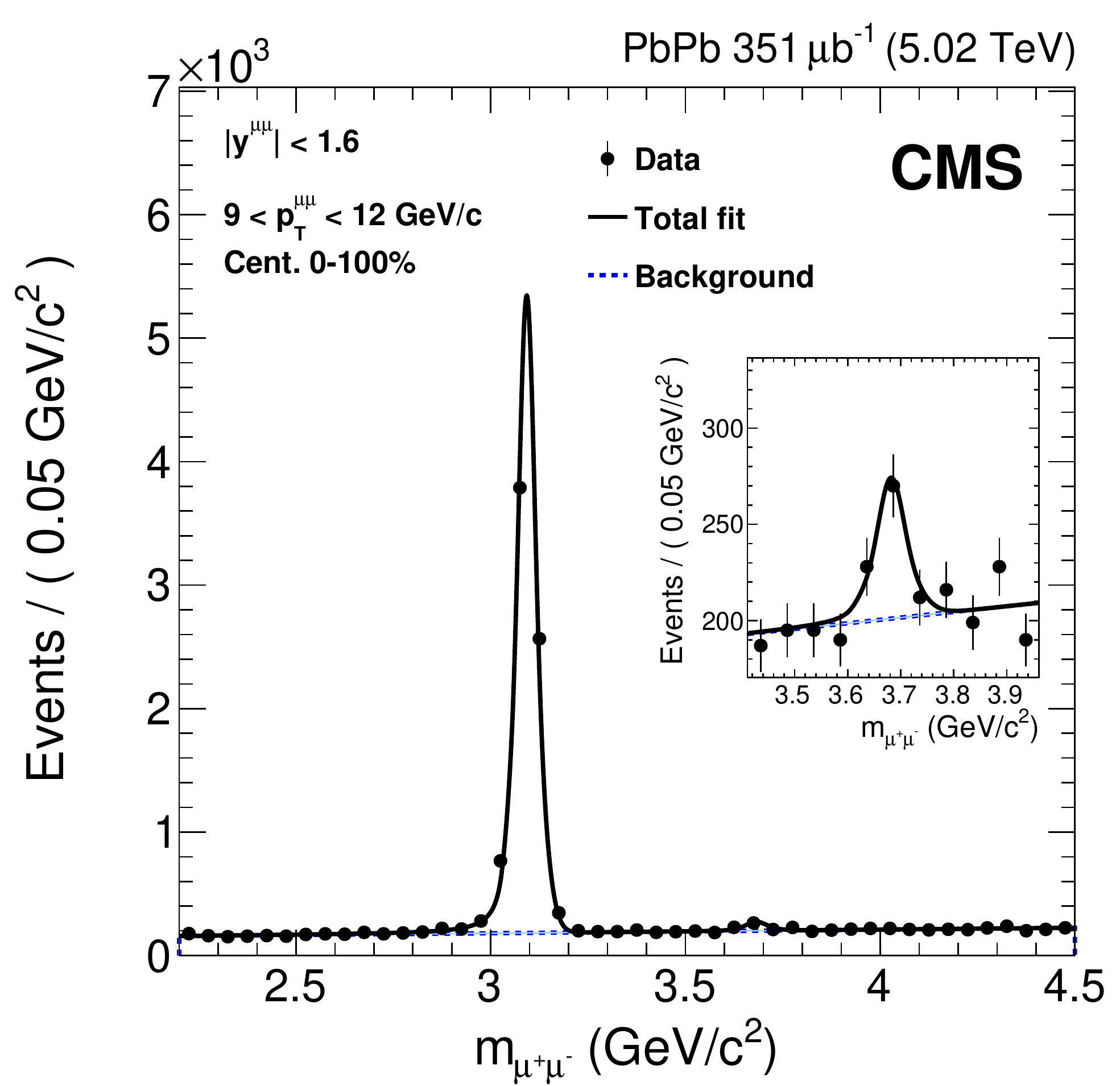}
        \includegraphics[width=0.49\textwidth]{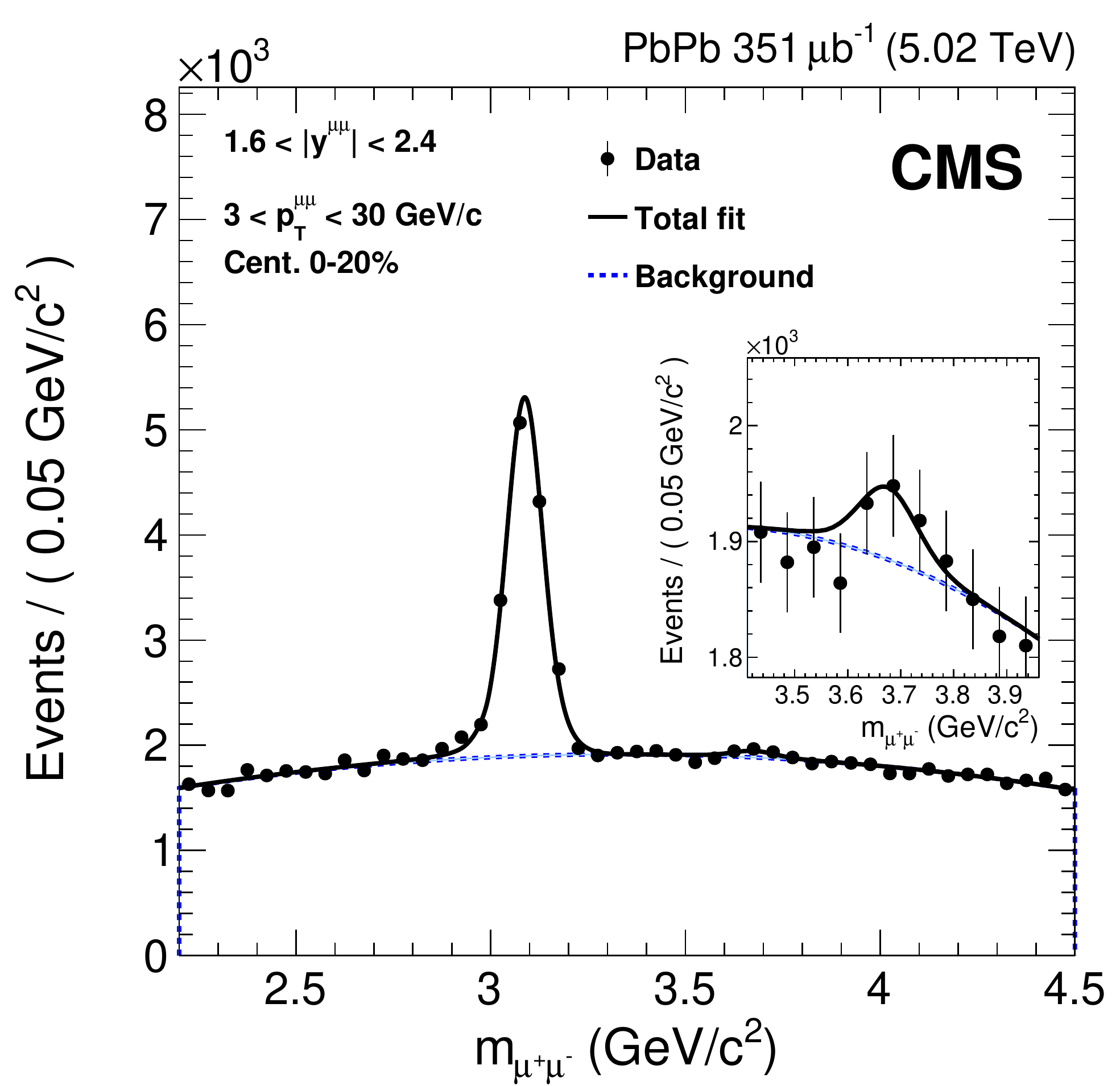}
    \caption{Invariant mass spectrum of \mumu pairs  (restricting to the \psiP region in the insets) in \PbPb collisions for (\cmsLeft) $\abs{y}<1.6$, $9<\pt<12\GeVc$, all centrality, and (\cmsRight) $1.6<\abs{y}<2.4$, $3 < \pt < 30\GeVc$, 0--20\% centrality. The results of the fits described in the text are also shown.
      }
    \label{fig:invmass_pbpb}
\end{figure}

Integrated over centrality, rapidity, and \pt, the fits yield about 38\,000 (293\,000) \JPsi and 530 (11\,200) \psiP mesons in \PbPb (\pp)  collisions. Examples of such fits for the \PbPb data are shown in Fig.~\ref{fig:invmass_pbpb}, for two cases of very different \psiP signal-to-background ratios.

The systematic uncertainties arise from the signal and background fitting model assumptions, the imperfect efficiency cancellation, and the nonprompt
residual contamination. These uncertainties are derived separately for \pp and \PbPb data, and the total systematic uncertainty is computed as the quadratic sum of the partial terms.

In order to determine the uncertainty associated with the fitting procedure, the signal and background models are independently varied in each analysis bin. For the signal, the fixed parameters are released one by one. As a further test, the signal parameters are fixed to the values obtained from a \psiP simulation, instead of the \JPsi simulation. A different signal shape is also tried: a CB function plus a Gaussian function. For the background model,
the fitted mass range is varied and
an exponential of a polynomial is used, redoing LLR tests to choose the best order for the polynomial in each analysis bin. The maximum difference of the single ratio $N(\psiP)/N(\JPsi)$ between the nominal and alternative fits, performed for signal and background separately, is taken as the corresponding systematic uncertainty. These uncertainties depend crucially on the signal-to-background ratio in the \psiP region. The absolute uncertainties on the double ratio remain below 0.02 and 0.11 for the \pp and \PbPb contributions, respectively.

The nonprompt \JPsi and \psiP fractions in \pp collisions, as well as the \JPsi fraction in \PbPb
collisions, are validated
with two-dimensional fits to the dimuon mass and pseudo-proper decay length distributions~\cite{Chatrchyan:2012np}.
The \PbPb event sample does not have enough \psiP events to provide a reliable two-dimensional fit. The variation in the double ratio when using nonprompt fractions from the two-dimensional fits is taken as a systematic uncertainty, never exceeding 0.07.

Finally, residual noncancellations of efficiencies in the double ratio are evaluated with MC studies, considering a broad range of \pt spectra compatible with the \pp and \PbPb data within their uncertainties. The corresponding systematic uncertainty varies between 0.01 and 0.05, with the exception of the lowest \pt bin, where it reaches 0.10. If the quarkonium acceptances were different in \pp and \PbPb, they would not perfectly cancel in the double ratio. This would be the case if some physics
effects
(such as polarization or energy loss) would affect quarkonia in \PbPb collisions with a strong kinematic dependence within an analysis bin. As in previous analyses~\cite{Chatrchyan:2011pe,Chatrchyan:2012lxa,Chatrchyan:2013nza,hin-12-007}, such possible effects are considered as part of the physics
under study and not as systematic uncertainties.

\begin{figure}[htbp]
 \centering
 \includegraphics[width=0.49\textwidth]{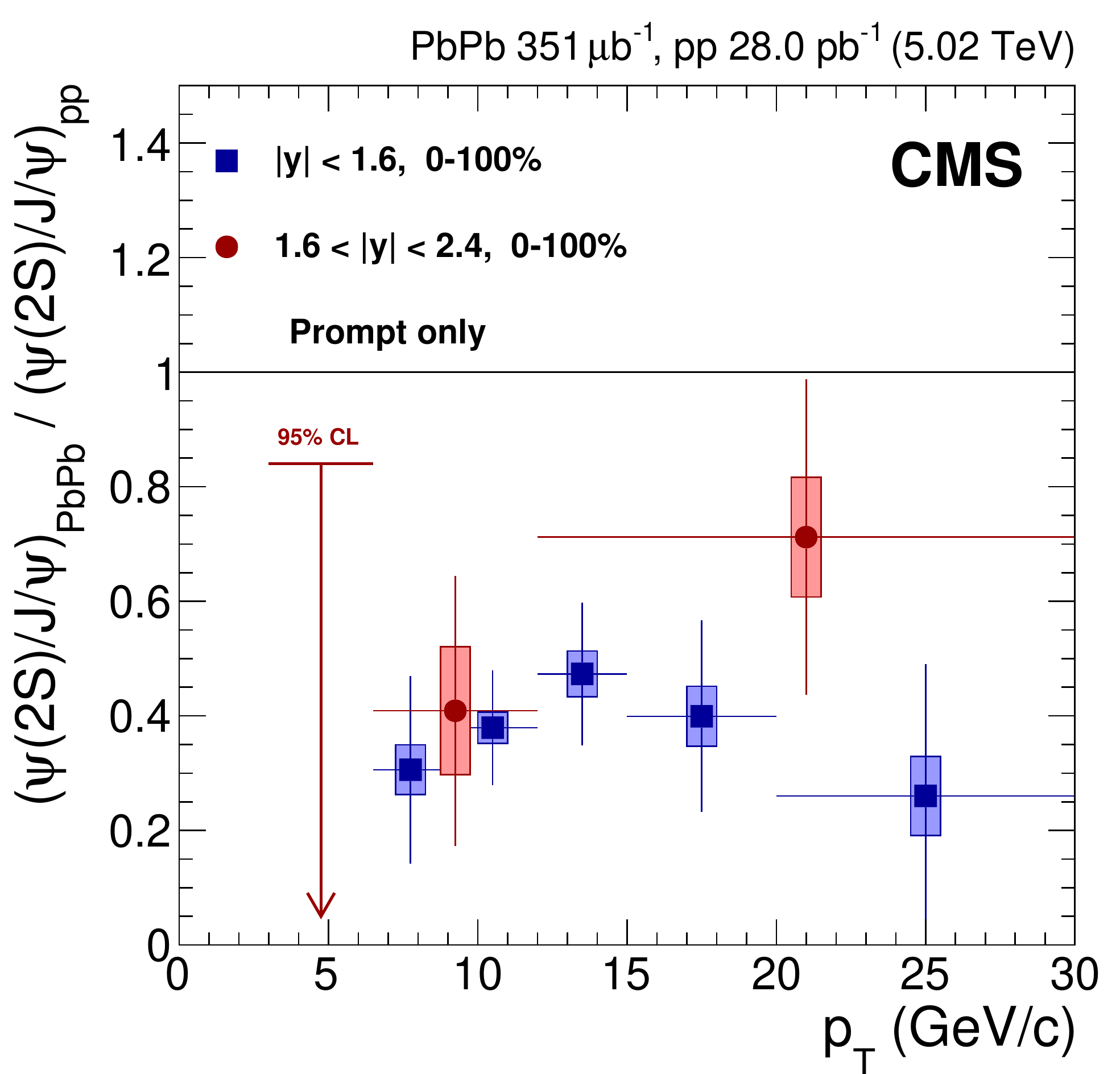}
 \caption{Transverse momentum dependence of \doubleRatio, for mid (squares) and forward (circles) rapidity, with both muons above the \pt threshold described in the text.
The arrow represents the 95\% \CL interval in the bin where the measurement is consistent with 0.
The vertical lines (boxes) represent the statistical (systematic) uncertainties.
The horizontal lines represent the width of the \pt bins.
 \label{fig:result_pt}}
\end{figure}

\begin{figure}[htbp]
 \centering
 \includegraphics[width=0.49\textwidth]{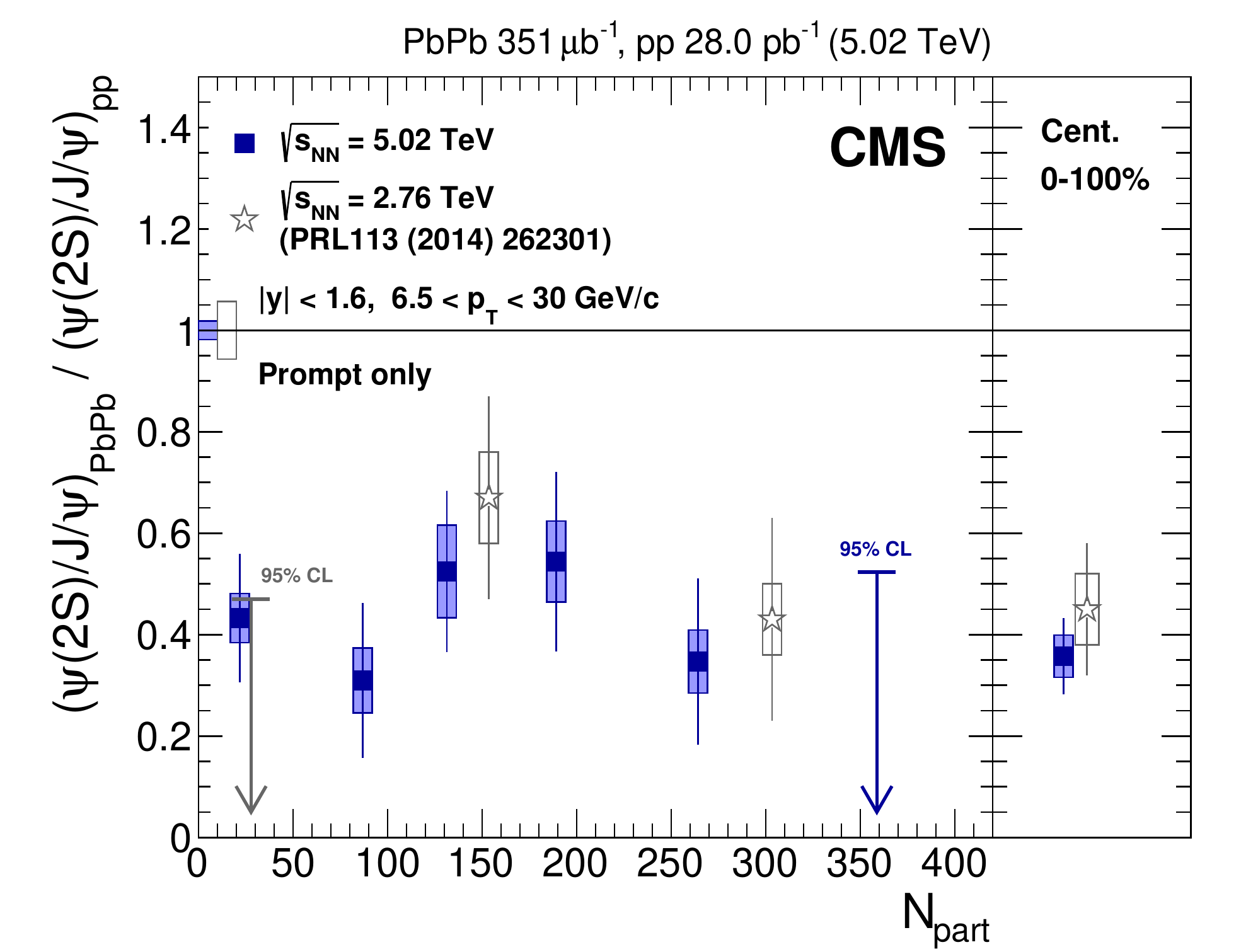}
 \includegraphics[width=0.49\textwidth]{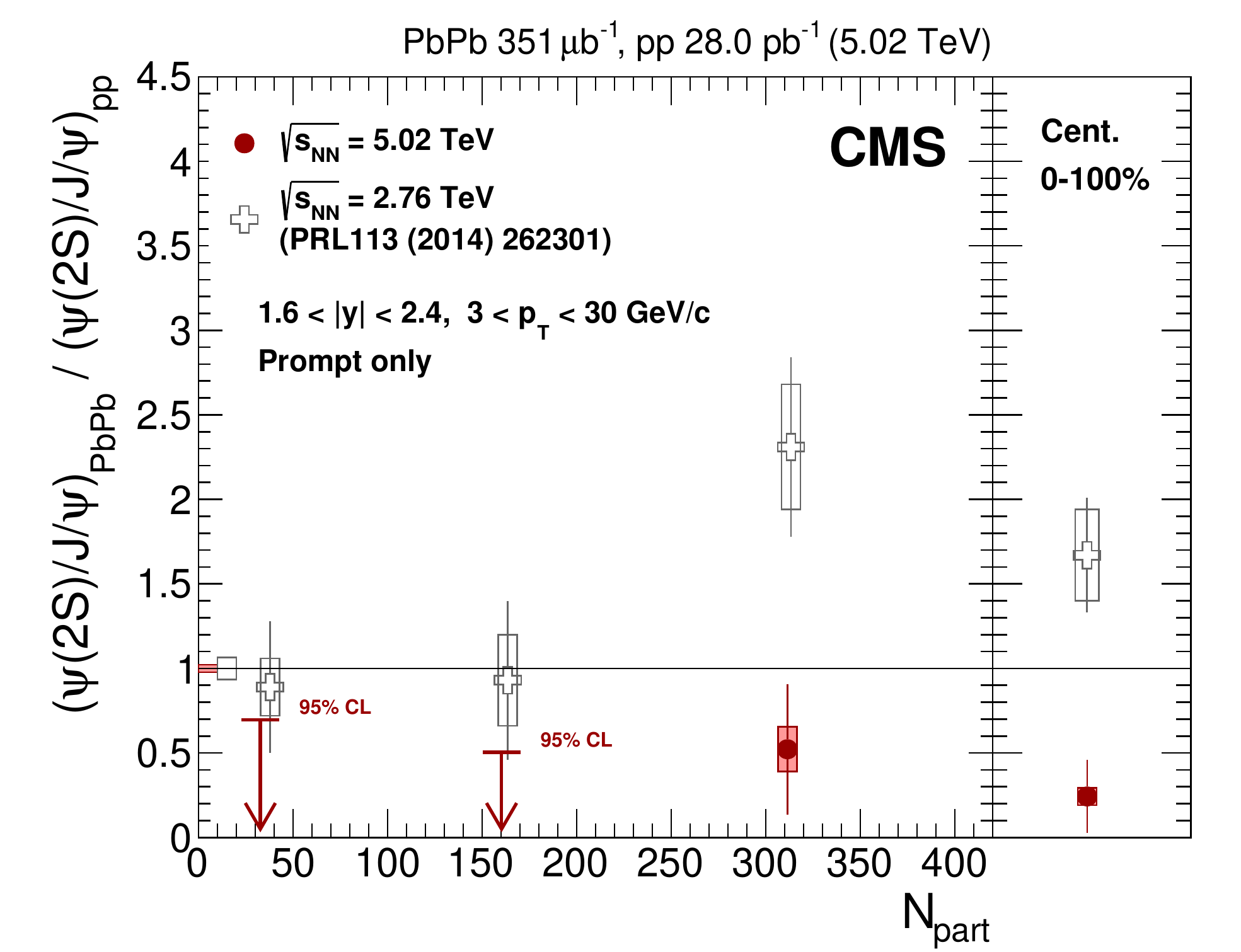}
 \caption{
    Event centrality dependence of \doubleRatio, for mid (\cmsLeft)
    and forward
    (\cmsRight)
    rapidity, with both muons above the \pt threshold described in the text. Values for the centrality-integrated sample are given in the right panels.
    The arrows represent 95\% \CL intervals in the bins where the measurement is consistent with~0.
    The vertical lines (boxes) represent the statistical (systematic) uncertainties. The statistical and systematic uncertainties in the \pp measurements, common to all points, are represented as boxes at unity. The measurements from CMS at $\sqrtsnn = 2.76\TeV$~\cite{hin-12-007} are also shown.
 \label{fig:result_cent}}
\end{figure}

The measured double ratio is shown in Figs.~\ref{fig:result_pt} and~\ref{fig:result_cent} as a function of \pt and event centrality, respectively. Centrality is commonly represented by the average number of participating nucleons, $\langle N_\text{part} \rangle$,
computed with the Glauber model~\cite{Miller:2007ri}. In terms of centrality percentiles, the bins correspond to 0--10, 10--20, 20--30, 30--40, 40--50, and 50--100\% in the midrapidity region, and 0--20, 20--40, and 40--100\% for the forward rapidity region. The most ``peripheral'' bins are rather wide and, since quarkonium yields scale with the number of nucleon-nucleon collisions, most charmonia are produced close to the most central edge of the bins. The $\langle N_\text{part} \rangle$ values used in the following are computed for events following a flat centrality distribution. When the measured double ratio is consistent with zero within one standard deviation of its statistical uncertainty, its corresponding 95\% confidence level (CL) interval is computed, using the Feldman--Cousins procedure~\cite{Feldman:1997qc}.
The numerical values of all measurements, including the 95\% \CL intervals, are tabulated in \suppMaterial.

The rightmost panels in Fig.~\ref{fig:result_cent} show the double ratio integrated over \pt and centrality: $0.36 \pm 0.08\stat \pm0.05\syst$ in the $\abs{y}<1.6$ and $6.5 < \pt < 30\GeVc$ range, and $0.24 \pm 0.22\stat \pm 0.09\syst$ in the $1.6<\abs{y}<2.4$ and $3 < \pt < 30 \GeVc$ range.

The double ratios measured at 5.02\TeV and reported in this paper are below unity in all bins.
Assuming that the \JPsi is suppressed in \PbPb collisions at $\sqrtsnn = 5.02\TeV$, as suggested by results at lower energy in the same kinematic range by CMS~\cite{Chatrchyan:2012np} or at both energies but in a different rapidity range by ALICE~\cite{Abelev:2013ila,Adam:2016rdg}, the
\psiP is more suppressed than the \JPsi in \PbPb collisions. This difference in suppression is already present in the most peripheral ranges probed by this analysis, starting at 40 or 50\% centrality. No strong dependencies are observed with centrality or transverse momentum.

In Fig.~\ref{fig:result_cent}, a reasonable agreement with the measurement made at $\sqrtsnn = 2.76\TeV$ can be seen in most of the bins. Systematic uncertainties are uncorrelated between the two datasets. 
In the range $1.6<\abs{y}<2.4$ and $3 < \pt < 30 \GeVc$, the double ratios are consistently lower in the 5.02\TeV data, especially in the most central collisions.
The difference is at the level of around 3 standard deviations in the centrality-integrated sample.

In summary, the double ratio \doubleRatio was measured to compare the relative production of \JPsi and \psiP mesons in \pp and \PbPb collisions at $\sqrtsnn = 5.02\TeV$, as a function of transverse momentum and collision centrality. The double ratio is below unity in all bins, suggesting that the \psiP
yield is more suppressed than the \JPsi yield
in the kinematic range explored.
The 5.02\TeV data do not show the enhancement in the double ratio previously seen for collisions at 2.76\TeV in the $1.6 < \abs{y} < 2.4$ and $3 < \pt < 30\GeVc$ range.
No strong variations are observed with charmonium \pt or collision centrality.
These results should significantly contribute to a deeper understanding of the medium effects at play in \JPsi and \psiP production, in particular by better constraining the energy dependence of the regeneration effects potentially affecting the two charmonium states.

\hyphenation{Bundes-ministerium Forschungs-gemeinschaft Forschungs-zentren} We congratulate our colleagues in the CERN accelerator departments for the excellent performance of the LHC and thank the technical and administrative staffs at CERN and at other CMS institutes for their contributions to the success of the CMS effort. In addition, we gratefully acknowledge the computing centers and personnel of the Worldwide LHC Computing Grid for delivering so effectively the computing infrastructure essential to our analyses. Finally, we acknowledge the enduring support for the construction and operation of the LHC and the CMS detector provided by the following funding agencies: BMWFW and FWF (Austria); FNRS and FWO (Belgium); CNPq, CAPES, FAPERJ, and FAPESP (Brazil); MES (Bulgaria); CERN; CAS, MoST, and NSFC (China); COLCIENCIAS (Colombia); MSES and CSF (Croatia); RPF (Cyprus); SENESCYT (Ecuador); MoER, ERC IUT, and ERDF (Estonia); Academy of Finland, MEC, and HIP (Finland); CEA and CNRS/IN2P3 (France); BMBF, DFG, and HGF (Germany); GSRT (Greece); OTKA and NIH (Hungary); DAE and DST (India); IPM (Iran); SFI (Ireland); INFN (Italy); MSIP and NRF (Republic of Korea); LAS (Lithuania); MOE and UM (Malaysia); BUAP, CINVESTAV, CONACYT, LNS, SEP, and UASLP-FAI (Mexico); MBIE (New Zealand); PAEC (Pakistan); MSHE and NSC (Poland); FCT (Portugal); JINR (Dubna); MON, RosAtom, RAS, and RFBR (Russia); MESTD (Serbia); SEIDI and CPAN (Spain); Swiss Funding Agencies (Switzerland); MST (Taipei); ThEPCenter, IPST, STAR, and NSTDA (Thailand); TUBITAK and TAEK (Turkey); NASU and SFFR (Ukraine); STFC (United Kingdom); DOE and NSF (USA).
\bibliography{auto_generated}
\ifthenelse{\boolean{cms@external}}{}{
\clearpage
\appendix
\section{Supplementary information\label{app:suppMat}}
The numerical values of all measurements, including the 95\% \CL intervals, are summarized in Tables~\ref{tab:result_pt} and~\ref{tab:result_cent}.
\newcolumntype{X}{D{,}{\text{--}}{3.3}}
\begin{table*}[htb]
\topcaption{Transverse momentum dependence of the double ratio \doubleRatio, for mid-rapidity ($\abs{y}<1.6$) and forward rapidity ($1.6<\abs{y}<2.4$), with both muons within the CMS acceptance. The 95\% \CL intervals are also given.
 \label{tab:result_pt}}
\centering
\begin{scotch}{Xcr}
\multicolumn{1}{c}{\pt (\GeVcns{})} & \multicolumn{1}{c}{\doubleRatio} & 95\% \CL interval\\
\hline
\\
\multicolumn{3}{c}{$\abs{y}<1.6$}\\[1ex]

6.5,9 & $0.31 \pm 0.16\stat\pm 0.04\syst$ & $[0.01,0.65]$\\
9,12 & $0.38 \pm 0.10\stat\pm 0.03\syst$ & $[0.18,0.59]$\\
12,15 & $0.47 \pm 0.13\stat\pm 0.04\syst$ & $[0.23,0.75]$\\
15,20 & $0.40 \pm 0.17\stat\pm 0.05\syst$ & $[0.08,0.78]$\\
20,30 & $0.26 \pm 0.23\stat\pm 0.07\syst$ & $[0,0.81]$\\
\\
\multicolumn{3}{c}{$1.6<\abs{y}<2.4$}\\[1ex]

3,6.5 & $0.15 \pm 0.37\stat\pm 0.06\syst$ & $[0,0.84]$\\
6.5,12 & $0.41 \pm 0.24\stat\pm 0.11\syst$ & $[0,0.89]$\\
12,30 & $0.71 \pm 0.28\stat\pm 0.11\syst$ & $[0.17,1.34]$\\
\end{scotch}
\end{table*}

\begin{table*}[bht]
\topcaption{Event centrality dependence of the double ratio \doubleRatio, for
two kinematic ranges,
with both muons within the CMS acceptance. The 95\% \CL intervals are also given.
 \label{tab:result_cent}}
\newcolumntype{x}{D{,}{\,\pm\,}{3.1}}
\centering
\begin{scotch}{Xxrr}
\multicolumn{1}{c}{Centrality} & \npart & \doubleRatio & 95\% \CL interval\\
\hline
\\
\multicolumn{4}{c}{$\abs{y}<1.6$, $6.5<\pt<30\GeVc $}\\[1ex]
0,100\% & 114,8 & $0.36 \pm 0.08\stat\pm 0.05\syst$ & $[0.18,0.54]$\\[1ex]
0,10\% & 359,2 & $0.14 \pm 0.17\stat\pm 0.11\syst$ & $[0,0.55]$\\
10,20\% & 264,2 & $0.35 \pm 0.16\stat\pm 0.07\syst$ & $[0.01,0.70]$\\
20,30\% & 189,4 & $0.54 \pm 0.18\stat\pm 0.08\syst$ & $[0.17,0.93]$\\
30,40\% & 131,4 & $0.53 \pm 0.16\stat\pm 0.09\syst$ & $[0.18,0.89]$\\
40,50\% & 87,3 & $0.31 \pm 0.15\stat\pm 0.08\syst$ & $[0,0.58]$\\
50,100\% & 22,1 & $0.43 \pm 0.13\stat\pm 0.06\syst$ & $[0.17,0.72]$\\
\\
\multicolumn{4}{c}{$1.6<\abs{y}<2.4$, $3<\pt<30\GeVc $}\\[1ex]
0,100\% & 114,8 & $0.24 \pm 0.22\stat\pm 0.09\syst$ & $[0,0.67]$\\[1ex]
0,20\% & 312,2 & $0.52 \pm 0.39\stat\pm 0.15\syst$ & $[0,1.30]$\\
20,40\% & 160,3 & $-0.14 \pm 0.34\stat\pm 0.10\syst$ & $[0,0.52]$\\
40,100\% & 33,3 & $0.22 \pm 0.25\stat\pm 0.10\syst$ & $[0,0.72]$\\
\end{scotch}
\end{table*}

}

\cleardoublepage \section{The CMS Collaboration \label{app:collab}}\begin{sloppypar}\hyphenpenalty=5000\widowpenalty=500\clubpenalty=5000\textbf{Yerevan Physics Institute,  Yerevan,  Armenia}\\*[0pt]
A.M.~Sirunyan, A.~Tumasyan
\vskip\cmsinstskip
\textbf{Institut f\"{u}r Hochenergiephysik,  Wien,  Austria}\\*[0pt]
W.~Adam, E.~Asilar, T.~Bergauer, J.~Brandstetter, E.~Brondolin, M.~Dragicevic, J.~Er\"{o}, M.~Flechl, M.~Friedl, R.~Fr\"{u}hwirth\cmsAuthorMark{1}, V.M.~Ghete, C.~Hartl, N.~H\"{o}rmann, J.~Hrubec, M.~Jeitler\cmsAuthorMark{1}, A.~K\"{o}nig, I.~Kr\"{a}tschmer, D.~Liko, T.~Matsushita, I.~Mikulec, D.~Rabady, N.~Rad, B.~Rahbaran, H.~Rohringer, J.~Schieck\cmsAuthorMark{1}, J.~Strauss, W.~Waltenberger, C.-E.~Wulz\cmsAuthorMark{1}
% ...
\vskip\cmsinstskip
% ...
1:~~Also at Vienna University of Technology, Vienna, Austria\\
% ...

\end{sloppypar}
\end{document}